# Hybrid Analog-Digital Beamforming for Multiuser MIMO Millimeter Wave Relay Systems

Xuan Xue[+], Tadilo Endeshaw Bogale[++], Xianbin Wang[++], Yongchao Wang[+] and Long Bao Le[*]
Xidian University, China[+]
University of Western Ontario, Canada[++]
Institute National de la Recherche Scientifique (INRS), Canada[*]
Email: {xxue4, tadilo.bogale, xianbin.wang}@uwo.ca, ychwang@mail.xidian.edu.cn, long.le@emt.inrs.ca

*Abstract*— This paper proposed new hybrid, analog-digital, beamforming for a multiuser millimeter wave (mm-wave) relay system. For this system, we consider a sum rate maximization problem. The proposed hybrid beamforming is designed indirectly by considering a sum mean square error (MSE) minimization problem while utilizing the solution of digital beamforming. To this end, we assume that the digital beamforming utilizes the well known block diagonalization (BD) approach. Under this assumption, we solve our problem as follows: First, we formulate the sum rate maximization problem as the minimization of the MSE between the received signal of the hybrid and digital beamforming designs. Then, we design the hybrid beamformings of the source, relay and each destination by leveraging compressive sensing techniques. Simulation results confirm that the proposed hybrid beamforming design achieves performance very close to that of the digital one. Furthermore, we have examined the effects of the number of radio frequency (RF) chains and paths together, and the accuracy of angle of arrival (AoA) and angle of departure (AoD) estimators on the sum rate of the hybrid beamforming designs. Computer simulations reveal that the total sum rate of the hybrid beamforming increases when the number of RF chains and paths increase (or the accuracy of the AoA (AoD) estimator improves).

## I. INTRODUCTION

Despite the research efforts to deploy efficient wireless technologies, wireless industries always face spectrum crunch at microwave frequency bands (i.e., frequencies up to 6 GHz). Due to this reason, there is an interest to exploit the underutilized millimeter wave (mm-wave) frequencies (typical values $30 - 300$GHz) for cellular applications [1], [2]. However, the mm-wave frequencies face severe path loss, penetration loss and rain fading, and they are easily absorbed or scattered by gases [3]. Nevertheless, the small wavelength of mm-wave ensures that a large number of antennas can be deployed, which provide large antenna array beamforming gains to compensate for the severe propagation losses of mm-wave systems.

Even with significant beamforming gains, however, the non-line-of-sight (NLOS) communication is still a major obstacle for mm-wave frequencies as they have high propagation losses due to the effects of blockage [4]. To overcome the poor performance caused by the NLOS scenario and enable long distance mm-wave communication, relay assisted communications can be adopted to efficiently transmit the signals between the source and destination. With the help of a relay, the signals can be transmitted over a longer distance and the channels between source and relay (relay and destination) can be line-of-sight (LOS) which consequently help achieve a better system performance than the scenario where the channel between the source and destination is NLOS and transmission takes place without the help of relay. Furthermore, in contrast to the higher capital and operating cost of base stations, relay stations are comprised of simpler, less expensive, and lower operational costs. Specifically, this paper considers amplify-and-forward (AF) relays just for their simplicity. Furthermore, we assume that there is no direct link between the source and destination(s).

In a conventional multiantenna system, relay beamforming has already been adopted for microwave frequency applications due to its ability to improve the capacity and coverage area of wireless networks [5]–[7]. There are a bunch of beamforming approaches developed for relay systems in the past couple of decades [8]–[10]. However, these approaches are designed mainly for systems with a small number of antennas (around $\sim 10$) and employ the conventional digital beamforming. Systems based on digital beamforming requires the same number of radio frequency (RF) chains as that of the number of antennas where each RF chain requires extra circuit and power consumption. Thus, when the number of antennas is very large (e.g., in mm-wwave systems), the deployment of digital beamforming scheme will be practically infeasible. For this reason, it is interesting to realize beamforming with a limited number of RF chains. One approach of achieving this goal is to deploy beamforming at both the digital and analog domains, i.e., hybrid beamforming. In the digital domain, beamforming is realized using microprocessors whereas, in the analog domain beamforming is implemented by employing low cost phase shifters [4], [11].

A number of hybrid beamforming designs are proposed in the past few years. In [2], [12], [13], a hybrid beamforming architecture is suggested for single user massive MIMO systems where matching pursuit (MP) algorithm is utilized [14]. In [15], hybrid precoding scheme for multiuser massive multiple input multiple output (MIMO) systems is considered. The paper employs the zero forcing (ZF) hybrid precoding where it is designed to maximize the sum rate of all users. In [1], a beam alignment technique using adaptive subspace sampling and hierarchical beam codebooks is proposed for mm-wave cellular networks. In [16], the sum rate maximization problem for the downlink massive MIMO systems is studied using the MP solution approach. In [17], a beam training (or beam steering) problem for the 60 GHz mm-wave communications is formulated as a numerical optimization problem such that the received signal is maximized. In [18], a user scheduling design approach employing hybrid beamforming for massive MIMO systems is proposed. All the aforementioned hybrid beamforming designs are considered for conventional massive MIMO (mm-wave) systems without introducing relay. In [19], a hybrid beamforming algorithm is designed for single user mm-wave relay systems. This paper employs a modified version of the compressive sensing algorithms proposed in [2].

In the current paper, we extend the work of [19] to a multiuser mm-wave relay system. In this regard, we consider a sum rate maximization problem and utilize the modified version of the MP method which is a particular type of the compressive sensing algorithm [16]. The hybrid beamforming is designed indirectly by considering a sum mean square error minimization problem while utilizing the solution of digital beamforming like in [16]. In this regard, we assume that the digital beamforming utilizes the well known block diagonalization (BD) approach as in [20]–[22]. Under this assumption, we solve our problem as follows: First, we formulate the sum rate maximization problem as the minimization of the mean square error (MSE) between the received signal of the hybrid and digital beamforming designs. Then, we design the hybrid beamformings of the source, relay and each destination by modifying the MP approach used in [16].

As will be clear in Section II, the beamforming matrix structures of the source and destinations are similar to those presented in [16]. Thus, the source and destination hybrid beamformings are designed by extending the compressive sensing approach of [16] to our system setup. However, the hybrid beamforming structure of the relay consists of two coupled analog beamforming matrices. For this reason, one can not utilize the compressive sensing approach of [16] (or [2]) directly to design the relay hybrid beamforming matrix. To this end, we propose a two step approach to design the hybrid beamforming matrices at the relay station which is detailed in Section V. Simulation results confirm that the proposed hybrid beamforming design achieves performance very close to that of the digital one. Furthermore, we have examined the effects of the number of RF chains and paths together, and the accuracy of angle of arrival (AoA) and angle of departure (AoD) estimators on the sum rate of the proposed hybrid beamforming designs. Numerical results reveal that the total sum rate of the hybrid beamforming increases when the number of RF chains and paths increase (or the accuracy of the AoA (AoD) estimator improves).

This paper is organized as follows. Sections II and III discuss the system and channel models, and conventional digital beamformings. In Section IV, the proposed hybrid beamforming design problem formulation is presented. And, Section V provides the proposed algorithm. In Section VI, simulation results are used to asses the performance of the proposed hybrid beamforming and to study the relation between the hybrid and digital beamformings. Conclusions are drawn in Section VII.

*Notations:* In this paper, upper/lower-case boldface letters denote matrices/column vectors. $\mathbf{X}(i,j)$, $||\mathbf{X}||_F$, $\text{trace}(\mathbf{X})$, $\mathbf{X}^T$, $\mathbf{X}^H$ and $\text{E}(\mathbf{X})$ denote the $(i,j)$th element, Frobenius norm, trace, transpose, conjugate transpose and expected value of $\mathbf{X}$, respectively. $\mathbf{I}_n$ is the identity matrix of size $n \times n$, and $\mathbb{C}^{M \times N}$ represent spaces of $M \times N$ matrices with complex values. The acronym null, s.t and i.i.d denote "(right) null space", "subject to" and "independent and identically distributed", respectively. The $\text{diag}(.)$ and $\text{blkdiag}(.)$ denote diagonal and block diagonal of matrices.

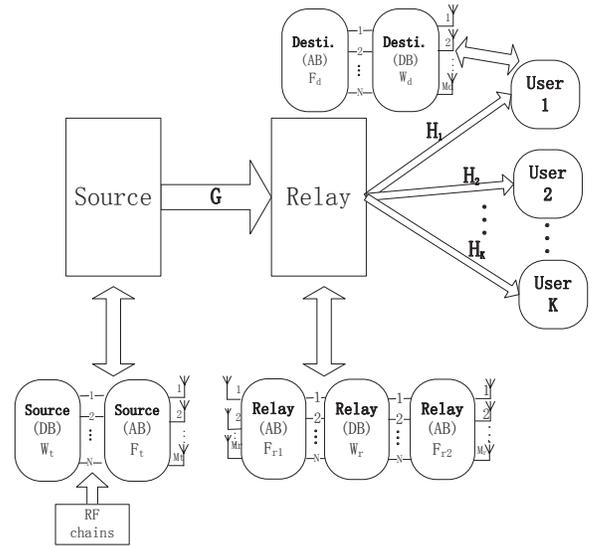

Fig. 1. Hybrid beamforming system

## II. SYSTEM MODELS

This section discusses the system model for the considered multiuser MIMO relay system. As mentioned in the introduction section, the proposed communication system employs a combination of both analog and digital architectures which is shown in Fig 1. As we can see from this figure, the source and relay stations are serving $K$ decentralized destinations where number of antennas (RF chains) at the source, relay and each of the destinations are $M_s$, $M_r$ and $M_d$ ($N_s$, $N_r$ and $N_d$ RF chains), respectively.

For convenience, let us represent the transmitted symbols of the $k$th destination as $\mathbf{s}_k \in \mathbb{C}^{S_k \times 1}$. Thus, the overall symbols transmitted from the source can be expressed as $\mathbf{s} = [\mathbf{s}_1^T, \mathbf{s}_2^T, \cdots, \mathbf{s}_K^T]^T$ with $S_K = \sum_{i=1}^{K} S_i$. Under such settings, the transmitted signal at source can be expressed as

$$\mathbf{x}_t = \mathbf{F}_t \mathbf{W}_t \mathbf{s} \qquad (1)$$

where $\mathbf{W}_t \in \mathbb{C}^{N_s \times S_K}$ and $\mathbf{F}_t \in \mathbb{C}^{M_s \times N_s}$ are the digital and analog beamforming matrices of the proposed architecture and $\mathbf{s}$ is assumed to have $\mathbb{E}[\mathbf{s}\mathbf{s}^H] = \frac{1}{S_K}\mathbf{I}_{S_K}$. Since the analog beamforming matrix $\mathbf{F}_t$ is realized using phase shifters, its

elements satisfy the fact that $|\mathbf{F}_t(i,j)|^2 = 1$. The total power constraint at the source is[1]

$$P_s = \mathbb{E}[\mathbf{x}_t^H \mathbf{x}_t] = \text{trace}(\mathbf{W}_t^H \mathbf{F}_t^H \mathbf{F}_t \mathbf{W}_t). \quad (2)$$

The received signal at the relay can be expressed as

$$\mathbf{y}_r = \mathbf{G} \mathbf{F}_t \mathbf{W}_t \mathbf{s} + \mathbf{n}_r \quad (3)$$

where $\mathbf{y}_r$ is the $M_r \times 1$ received vector, $\mathbf{G}$ is the $M_r \times M_s$ channel matrix between the source and relay station, and $\mathbf{n}_r$ is the addictive noise vector at relay where each entry of $\mathbf{n}_r$ is assumed to be i.i.d zero mean circularly symmetric complex Gaussian random variable each with variance $\sigma_r^2$ (i.e., $\mathbf{n}_r \sim \mathcal{CN}(\mathbf{0}, \sigma^2 \mathbf{I})$).

At the relay station, we may need to employ two analog beamforming matrices where one of them is used for the received signal from the source ($\mathbf{F}_{r2}$) and the other is used to beamform and forward the relay signal to the destinations ($\mathbf{F}_{r1}$), and one digital beamforming to perform baseband processing ($\mathbf{W}_r$). Using these beamforming matrices, the transmitted signal of the relay can be expressed as

$$\begin{aligned}\mathbf{x}_r &= \mathbf{F}_{r1} \mathbf{W}_r \mathbf{F}_{r2} \mathbf{y}_r \\ &= \mathbf{F}_{r1} \mathbf{W}_r \mathbf{F}_{r2} \mathbf{G} \mathbf{F}_t \mathbf{W}_t \mathbf{s} + \mathbf{F}_{r1} \mathbf{W}_r \mathbf{F}_{r2} \mathbf{n}_r \end{aligned} \quad (4)$$

where $\mathbf{F}_{r1} \in \mathbb{C}^{M_r \times N_r}$ and $\mathbf{F}_{r2} \in \mathbb{C}^{N_r \times M_r}$ are analog beamforming matrices, $\mathbf{W}_r$ is $N_r \times N_r$ digital beamforming matrix, and $\mathbf{x}_r$ is the $M_r \times 1$ transmitted signals at relay. The transmitted power at the relay is, therefore, given as

$$\begin{aligned}P_r &= \mathbb{E}[\mathbf{x}_r^H \mathbf{x}_r] \\ &= \text{trace}[\mathbf{F}_{r1} \mathbf{W}_r \mathbf{F}_{r2} (\mathbf{G} \mathbf{W}_t \mathbf{F}_t \mathbf{F}_t^H \mathbf{W}_t^H \mathbf{G}^H + \sigma_r^2 \mathbf{I}) \\ &\quad \times \mathbf{F}_{r2}^H \mathbf{W}_r^H \mathbf{F}_{r1}^H]. \end{aligned} \quad (5)$$

Similar to the source and relay stations, the $k$th destination also employs both analog beamforming $\mathbf{F}_{dk}$ and digital beamforming $\mathbf{W}_{dk}$. Using these beamforming matrices, the recovered signal at the $k$th destination is given as

$$\begin{aligned}\mathbf{y}_k &= \mathbf{W}_{dk}^H \mathbf{F}_{dk}^H \mathbf{H}_k^H \mathbf{x}_r + \mathbf{W}_{dk}^H \mathbf{F}_{dk}^H \mathbf{n}_{dk} \\ &= \mathbf{W}_{dk}^H \mathbf{F}_{dk}^H \mathbf{H}_k^H \mathbf{F}_{r1} \mathbf{W}_r \mathbf{F}_{r2} \mathbf{G} \mathbf{F}_t \mathbf{W}_t \mathbf{s} \\ &\quad + \mathbf{W}_{dk}^H \mathbf{F}_{dk}^H \mathbf{H}_k^H \mathbf{F}_{r1} \mathbf{W}_r \mathbf{F}_{r2} \mathbf{n}_r + \mathbf{W}_{dk}^H \mathbf{F}_{dk}^H \mathbf{n}_{dk} \end{aligned} \quad (6)$$

where $\mathbf{H}_k$ is the channel between the $k$th destination and relay station, and $\mathbf{F}_{dk} \in \mathbb{C}^{M_d \times N_d}$ ($\mathbf{W}_{dk} \in \mathbb{C}^{N_d \times M_d}$) is the analog (digital) beamforming matrix for the $k$th destination[2].

## III. CHANNEL MODEL AND DIGITAL BEAMFORMING

This section discusses the channel model used in the paper and the conventional digital beamforming.

---

[1] Here the elements of $|\mathbf{F}_t(i,j)|^2$ can have arbitrary constant value. In this paper, $|\mathbf{F}_t(i,j)|^2$ is set to one without loss of generality.

[2] As will be clear in the sequel, $\mathbf{F}_{r2}$ depends on $\mathbf{G}$ whereas, $\mathbf{F}_{r1}$ depends on the channels $\mathbf{H}_k, \forall k$. For these reasons, $\mathbf{F}_{r1}$ and $\mathbf{F}_{r2}$ may not necessarily correlate each other.

### A. Channel Model

Different from microwave wireless communication channels, mm-wave channels often have sparse structures which can be characterized by low rank matrices. This is due to the fact that such systems exhibit limited number of paths [2], [16]. As can be seen from Fig. 1, the considered system consists of the channels: $\mathbf{G}$ and $\mathbf{H}_k$.

The structures of $\mathbf{G}$ and $\mathbf{H}_k$ depend on the antenna array manifolds of the source, relay and destinations. If we deploy uniform linear (planar) arrays, we can represent these channels as

$$\begin{aligned}\mathbf{G} &= \sum_{l=1}^{L_g} \alpha_{lg} \mathbf{a}_{r,l}(\theta_{r,l}) \mathbf{a}_{s,l}^H(\theta_{s,l}), \\ \mathbf{H}_k &= \sum_{l=1}^{L_{hk}} \alpha_{lhk} \mathbf{a}_{dk,l}(\beta_{dk,l}) \mathbf{a}_{r,l}^H(\beta_{r,l}), \end{aligned} \quad (7)$$

where $L_g(L_{hk})$ is the number of propagation paths in $\mathbf{G}(\mathbf{H}_k)$, $\alpha_{lg}(\alpha_{lhk})$ is the propagation loss of the $l$th path in $\mathbf{G}(\mathbf{H}_k)$, $\mathbf{a}_{r,l}(\theta_{r,l})/\mathbf{a}_{s,l}(\theta_{s,l})$ is the relay/source array response vector for the channel $\mathbf{G}$, and $\mathbf{a}_{dk,l}(\beta_{dk,l})/\mathbf{a}_{r,l}(\beta_{r,l})$ is the $k$th destination/relay array response vector for the channel $\mathbf{H}_k$. Specifically, the current paper considers uniform linear arrays (ULA) where the array responses can be expressed as

$$\begin{aligned}\mathbf{a}_s(\theta) &= \frac{1}{\sqrt{M_s}}[1, \exp^{j\frac{2\pi}{\lambda}d\sin(\theta)}, \cdots, \exp^{j(M_s-1)\frac{2\pi}{\lambda}d\sin(\theta)}]^T, \\ \mathbf{a}_r(\theta) &= \frac{1}{\sqrt{M_r}}[1, \exp^{j\frac{2\pi}{\lambda}d\sin(\theta)}, \cdots, \exp^{j(M_r-1)\frac{2\pi}{\lambda}d\sin(\theta)}]^T, \\ \mathbf{a}_{dk}(\theta) &= \frac{1}{\sqrt{M_d}}[1, \exp^{j\frac{2\pi}{\lambda}d\sin(\theta)}, \cdots, \exp^{j(M_d-1)\frac{2\pi}{\lambda}d\sin(\theta)}]^T, \end{aligned} \quad (8)$$

where $j = \sqrt{-1}$, $\lambda$ is the transmission wave length and $d$ is the antenna spacing.

### B. Digital Beamforming

For better exposition of the proceeding sections, we summarize the conventional digital beamforming designs for relay networks. As discussed in the introduction section, the proposed hybrid beamforming design employs the solution of the digital beamformer. Although there are several existing digital beamforming algorithms, this paper adopts the solution obtained by the well-known BD beamforming scheme proposed in [20] since it is a low complexity algorithm which is summarized as follows:

For the conventional digital beamforming relay system, we assume the same number of antennas at different destinations as the one discussed in the previous section. Under such settings, the received signal at the $k$th destination can be expressed as

$$\mathbf{y}_k^D = \tilde{\mathbf{W}}_{dk}^H \mathbf{H}_k^H \tilde{\mathbf{W}}_r \mathbf{G} \tilde{\mathbf{W}}_t \mathbf{s} + \tilde{\mathbf{W}}_{dk}^H \mathbf{H}_k^H \tilde{\mathbf{W}}_r \mathbf{n}_r + \tilde{\mathbf{W}}_{dk}^H \mathbf{n}_{dk} \quad (9)$$

where $\tilde{\mathbf{W}}_r$, $\tilde{\mathbf{W}}_t$ and $\tilde{\mathbf{W}}_{dk}$ are the digital beamforming at relay, source and each destination, respectively. Now if we employ the BD beamforming approach of [20], the beamforming matrices are designed by following two steps.

In the first step, the channel between the source and relay is diagonalized by setting $\tilde{\mathbf{W}}_t = \mathbf{V}_g$ and $\tilde{\mathbf{W}}_r = \tilde{\mathbf{W}}\mathbf{U}_g^H$, where $\mathbf{V}_g$ and $\mathbf{U}_g$ are taken from the singular value decomposition (SVD) of the channel $\mathbf{G} = \mathbf{U}\Lambda\mathbf{V}^H$, where the diagonal elements of $\Lambda$ is arranged in decreasing order and $\mathbf{U}_g$ and $\mathbf{V}_g$ are taken from the first $S_K$ columns of $\mathbf{U}$ and $\mathbf{V}$. By doing so and after some mathematical manipulations, we can rewrite $\mathbf{y}_k^D$ as

$$\begin{aligned}
\mathbf{y}_k^D &= \tilde{\mathbf{W}}_{dk}^H (\mathbf{H}_k^H \mathbf{x}_r^D + \mathbf{n}_{dk}) \\
&= \tilde{\mathbf{W}}_{dk}^H (\mathbf{H}_k^H \sum_{i=1}^K \tilde{\mathbf{W}}_i(\tilde{\mathbf{s}}_i + \tilde{\mathbf{n}}_{ri}) + \mathbf{n}_{dk}) \\
&= \tilde{\mathbf{W}}_{dk}^H \mathbf{H}_k^H \tilde{\mathbf{W}}_k (\tilde{\mathbf{s}}_k + \tilde{\mathbf{n}}_{rk}) + \tilde{\mathbf{W}}_{dk}^H \mathbf{H}_k^H \sum_{i \neq k} \tilde{\mathbf{W}}_i(\tilde{\mathbf{s}}_i + \tilde{\mathbf{n}}_{ri}) \\
&\quad + \tilde{\mathbf{W}}_{dk}^H \mathbf{n}_{dk} \\
&= \mathbf{A}_{kk}\mathbf{s}_k + \sum_{i \neq k} \mathbf{A}_{ki}\mathbf{s}_i + \mathbf{B}_{kk}\tilde{\mathbf{n}}_{rk} + \sum_{i \neq k} \mathbf{B}_{ki}\tilde{\mathbf{n}}_{ri} + \tilde{\mathbf{W}}_{dk}^H \mathbf{n}_{dk}
\end{aligned} \quad (10)$$

where $\mathbf{n}_{dk} \sim (\mathbf{0}, \sigma_d^2 \mathbf{I})$, $\mathbf{A}_{kk} = \tilde{\mathbf{W}}_{dk}^H \mathbf{H}_k^H \tilde{\mathbf{W}}_k \Lambda_{gk}$, $\Lambda_g = \text{diag}([\text{diag}(\Lambda_{g1}), \text{diag}(\Lambda_{g2}), \cdots, \text{diag}(\Lambda_{gK})])$, $\mathbf{B}_{kk} = \tilde{\mathbf{W}}_{dk}^H \mathbf{H}_k^H \tilde{\mathbf{W}}_k$, and

$$\mathbf{x}_r^D = \sum_{i=1}^K \tilde{\mathbf{W}}_i(\tilde{\mathbf{s}}_i + \tilde{\mathbf{n}}_{ri}) \quad (11)$$

with $\tilde{\mathbf{s}} = [\tilde{\mathbf{s}}_1^T, \tilde{\mathbf{s}}_2^T, \cdots, \tilde{\mathbf{s}}_K^T]^T = \Lambda_g \mathbf{s}$, $\tilde{\mathbf{W}} = \tilde{\mathbf{W}}_r \mathbf{U}_g = [\tilde{\mathbf{W}}_1, \tilde{\mathbf{W}}_2, \cdots, \tilde{\mathbf{W}}_K]$, and $\tilde{\mathbf{W}}_i \in \mathbb{C}^{M_r \times S_k}, i = 1, \cdots, K$. The combined received signals for all destinations can be expressed as:

$$\mathbf{y}^D = \mathbf{R}\mathbf{s} + \mathbf{Q}\tilde{\mathbf{n}}_r + \tilde{\mathbf{W}}_d \mathbf{n}_d \quad (12)$$

where $\mathbf{R} = [\mathbf{R}_{11}, \mathbf{R}_{22}, \cdots, \mathbf{R}_{KK}]$, $\mathbf{Q} = [\mathbf{Q}_{11}, \mathbf{Q}_{22}, \cdots, \mathbf{Q}_{KK}]$, and $\tilde{\mathbf{W}}_d = \text{blkdiag}(\tilde{\mathbf{W}}_{d1}, \tilde{\mathbf{W}}_{d2}, \cdots, \tilde{\mathbf{W}}_{dK})$.

In the second step, the conventional multiuser BD algorithm [21], [22] is employed on $\mathbf{y}_k^D$ to ensure $\mathbf{R}_{ki} = \mathbf{Q}_{ki} = \mathbf{0}, \forall i \neq k$.

## IV. PROBLEM FORMULATION

For the hybrid architecture discussed in Fig. 1, the sum rate maximization problem can be mathematically formulated as

$$\begin{aligned}
\min_{\mathbf{F}_t, \mathbf{W}_t, \mathbf{F}_{r1}, \mathbf{F}_{r2}, \mathbf{W}_r, \mathbf{W}_d, \mathbf{F}_d} & \sum_{k=1}^K \sum_{i=1}^{S_k} \log_2(1+\gamma_{ki}) \\
\text{s.t.} \quad & \text{trace}(\mathbf{W}_t^H \mathbf{F}_t^H \mathbf{F}_t \mathbf{W}_t) = P_s, \\
& \text{trace}[\mathbf{F}_{r1}\mathbf{W}_r \mathbf{F}_{r2}(\mathbf{G}\mathbf{W}_t\mathbf{F}_t\mathbf{F}_t^H \mathbf{W}_t^H \mathbf{G}^H \\
& \quad + \sigma_r^2 \mathbf{I})\mathbf{F}_{r2}^H \mathbf{W}_r^H \mathbf{F}_{r1}^H] = P_r, \\
& |\mathbf{F}_t(i,j)|^2 = 1, \quad |\mathbf{F}_{r1}(i,j)|^2 = 1, \\
& |\mathbf{F}_{r2}(i,j)|^2 = 1, \quad |\mathbf{F}_{dk}(i,j)|^2 = 1,
\end{aligned} \quad (13)$$

where $\gamma_{ki}$ is the signal to interference to noise ratio (SINR) of the $i$th symbol of the $k$th destination which is given as

$$\gamma_{ki} = \frac{a_{ki}^2}{\sum_{m=1}^K \sum_{j=1,(m,j)\neq(k,i)}^{S_m} a_{mj}^2 + \sigma_r^2 b_{ki}^2 + \sigma_d^2 c_{ki}^2} \quad (14)$$

where $a_{ki} = |\mathbf{w}_{dki}^H \mathbf{F}_{dk}^H \mathbf{H}_k^H \mathbf{F}_{r1} \mathbf{W}_{rk} \mathbf{F}_{r2} \mathbf{G} \mathbf{F}_t \mathbf{W}_t|$, $b_{ki} = |\mathbf{w}_{dki}^H \mathbf{F}_{dk}^H \mathbf{H}_k^H \mathbf{F}_{r1} \mathbf{W}_{rk} \mathbf{F}_{r2}|$, $c_{ki} = |\mathbf{w}_{dki}^H \mathbf{F}_{dk}^H|$, and $\mathbf{W}_{dk} = [\mathbf{w}_{dk1}, \mathbf{w}_{dk2}, \cdots, \mathbf{w}_{dks_k}]$.

As we can see, the problem (13) consists of non-convex objective and constraint functions. Thus, getting the global optimal solution of this problem is not trivial. Furthermore, as the solution requires each of the elements of the analog beamforming matrices to be unity, examining the above problem using the standard optimization tools such as iterative Lagrangian multiplier method is almost infeasible. This is due to the fact that the number of introduced lagrangian multipliers will be extremely large as the dimensions of $\mathbf{F}_t, \mathbf{W}_t, \mathbf{F}_{r1}, \mathbf{F}_{r2}$ are large for mm-wave frequency applications. In the following, we reformulate the above problem such that the reformulated problem can be solved using low complexity algorithm.

In fact, one can notice from (6) and (10) that $\mathbf{y}_k$ and $\mathbf{y}_k^D$ have similar structures. Thus, one approach of examining the above problem can be achieved by choosing the hybrid beamforming matrices such that $\mathbf{y}_k$ is close to $\mathbf{y}_k^D$. In this regard, we compute the MSE between $\mathbf{y}_k$ and $\mathbf{y}_k^D$ as

$$\begin{aligned}
e_k &= \mathbb{E}[(\mathbf{y}_k - \mathbf{y}_k^D)^H(\mathbf{y}_k - \mathbf{y}_k^D)] \\
&= ||\mathbf{W}_{dk}^H \mathbf{F}_{dk}^H \mathbf{H}_k^H \mathbf{F}_{r1} \mathbf{W}_r \mathbf{F}_{r2} \mathbf{G} \mathbf{F}_t \mathbf{W}_t - \mathbf{A}_{1k}||_F^2 \\
&\quad + \sigma_r^2 ||\mathbf{W}_{dk}^H \mathbf{F}_{dk}^H \mathbf{H}_k^H \mathbf{F}_{r1} \mathbf{W}_r \mathbf{F}_{r2} - \mathbf{A}_{2k}||_F^2 \\
&\quad + \sigma_d^2 ||\mathbf{W}_{dk}^H \mathbf{F}_{dk}^H - \tilde{\mathbf{W}}_{dk}^H||_F^2
\end{aligned} \quad (15)$$

where $\mathbf{A}_{1k} = \tilde{\mathbf{W}}_{dk}^H \mathbf{H}_k^H \tilde{\mathbf{W}}_r \mathbf{G}\tilde{\mathbf{W}}_b$, $\mathbf{A}_{2k} = \tilde{\mathbf{W}}_{dk}^H \mathbf{H}_k^H \tilde{\mathbf{W}}_r$. The overall MSE of all destinations can be expressed as

$$\begin{aligned}
e_{MSE} &= \sum_{i=1}^K e_i = \mathbb{E}[(\mathbf{y} - \mathbf{y}^D)^H(\mathbf{y} - \mathbf{y}^D)] \\
&= ||\mathbf{W}_d^H \mathbf{F}_d^H \mathbf{H}^H \mathbf{F}_{r1} \mathbf{W}_r \mathbf{F}_{r2} \mathbf{G} \mathbf{F}_t \mathbf{W}_t - \mathbf{A}_1||_F^2 \\
&\quad + \sigma_r^2 ||\mathbf{W}_d^H \mathbf{F}_d^H \mathbf{H}^H \mathbf{F}_{r1} \mathbf{W}_r \mathbf{F}_{r2} - \mathbf{A}_2||_F^2 \\
&\quad + \sigma_d^2 ||\mathbf{W}_d^H \mathbf{F}_d^H - \tilde{\mathbf{W}}_d^H||_F^2
\end{aligned} \quad (16)$$

where $\mathbf{A}_1 = [\mathbf{A}_{11}, \mathbf{A}_{12}, \cdots, \mathbf{A}_{1K}]$, $\mathbf{A}_2 = [\mathbf{A}_{21}, \mathbf{A}_{22}, \cdots, \mathbf{A}_{2K}]$, $\mathbf{H} = [\mathbf{H}_1, \mathbf{H}_2, \cdots, \mathbf{H}_K]$, $\mathbf{W}_d = \text{blkdiag}(\mathbf{W}_{d1}, \mathbf{W}_{d2}, \cdots, \mathbf{W}_{dK})$ and $\mathbf{F}_d = \text{blkdiag}(\mathbf{F}_{d1}, \mathbf{F}_{d2}, \cdots, \mathbf{F}_{dK})$.

We aim to design the beamforming matrix $\mathbf{F}_t$, $\mathbf{W}_t$, $\mathbf{F}_{r1}$, $\mathbf{F}_{r2}$, $\mathbf{W}_r$, $\mathbf{W}_d$ and $\mathbf{F}_d$, which could minimize the MSE under power constraints at source and relay. Therefore, we optimize the hybrid beamforming matrices of (13) by examining the following problem

$$\begin{aligned}
\min_{\mathbf{F}_t, \mathbf{W}_t, \mathbf{F}_{r1}, \mathbf{F}_{r2}, \mathbf{W}_r, \mathbf{W}_d, \mathbf{F}_d} & \quad e_{MSE} \\
\text{s.t.} \quad & \text{trace}(\mathbf{W}_t^H \mathbf{F}_t^H \mathbf{F}_t \mathbf{W}_t) = P_t, \\
& \text{Tr}[\mathbf{F}_{r1}\mathbf{W}_r \mathbf{F}_{r2}(\mathbf{G}\mathbf{W}_t\mathbf{F}_t\mathbf{F}_t^H \mathbf{W}_t^H \mathbf{G}^H \\
& \quad + \sigma_r^2 \mathbf{I})\mathbf{F}_{r2}^H \mathbf{W}_r^H \mathbf{F}_{r1}^H] = P_r, \\
& |\mathbf{F}_t(i,j)|^2 = 1, \quad |\mathbf{F}_{r1}(i,j)|^2 = 1, \\
& |\mathbf{F}_{r2}(i,j)|^2 = 1, \quad |\mathbf{F}_d(i,j)|^2 = 1,
\end{aligned} \quad (17)$$

The above problem is still non-convex and hence global optimality cannot be guaranteed. However, such a problem can be solved using the MP algorithm leveraged from the compressive sensing domain which is detailed in the next section.

## V. PROPOSED ALGORITHM

Due to the fact that it is hard to solve problem (17) directly, we employ three steps to optimize the source, relay and destination beamforming matrices. In the first step, we optimize $\mathbf{W}_{dk}$ and $\mathbf{F}_{dk}$ for all destinations independently. In the second step, we optimize the relay beamforming matrices $\mathbf{F}_{r1}, \mathbf{F}_{r2}$, and $\mathbf{W}_r$ for the given $\mathbf{W}_d$ and $\mathbf{F}_d$ obtained from the first step. In the third step, the source beamforming matrices $\mathbf{F}_t$ and $\mathbf{W}_t$ are optimized by keeping $\mathbf{F}_{r1}, \mathbf{F}_{r2}, \mathbf{W}_r, \mathbf{W}_d, \mathbf{F}_d$ constant. Specifically, we apply the MP algorithm to obtain the beamforming matrices at the source and destinations, and we propose a two-step algorithm based on MP approach to get the hybrid beamforming matrices at the relay station. In order to explain the MP algorithm of the aforementioned steps clearly, let us consider the following general optimization problem:

$$\min_{\mathbf{F},\mathbf{W}} \quad ||\mathbf{D}_1\mathbf{F}\mathbf{W} - \mathbf{D}_2||_F^2$$
$$\text{s.t.} \quad \text{trace}(\mathbf{F}\mathbf{W}\mathbf{W}^H\mathbf{F}^H) = P,$$
$$|\mathbf{F}(i,j)|^2 = 1, \quad (18)$$

where $\mathbf{F}$ and $\mathbf{W}$ are analog and digital beamforming matrices. As we can see, this problem has the same mathematical structure as (18) of [16] which can be solved efficiently by using the MP algorithm as summarized in Table I. In this table, the codebook $\bar{\mathbf{A}}$ is selected from the estimated AoAs and AoDs [2], [16].

TABLE I
MATCHING PURSUIT ALGORITHM

---

**step 1.** Initialize: $\mathbf{F} = [\ ]$, $\mathbf{W}_{res} = \mathbf{W}_{opt} = \mathbf{D}_2$
**step 2.** For $i = 1 : N_{RF}$ do
  (1) $\mathbf{\Psi}_F = (\mathbf{D}_1\bar{\mathbf{A}})^H \mathbf{W}_{res}$;
  (2) $j = \arg\max(\text{diag}(\mathbf{\Psi}_F \mathbf{\Psi}_F^H))$;
  (3) $\mathbf{F} = [\mathbf{F}|\bar{\mathbf{A}}(:,j)]$;
  (4) $\hat{\mathbf{F}} = \mathbf{D}_1^H \mathbf{F}$;
  (5) $\mathbf{W} = (\hat{\mathbf{F}}^H \hat{\mathbf{F}})^{-1} \hat{\mathbf{F}}^H \mathbf{W}_{opt}$
  (6) $\mathbf{W}_{res} = \frac{\mathbf{W}_{opt} - \hat{\mathbf{F}}\mathbf{W}}{||\mathbf{W}_{opt} - \hat{\mathbf{F}}\mathbf{W}||_F}$;
  end
**step 3.** Scale $\mathbf{W} = \sqrt{P}\frac{\mathbf{W}}{||\mathbf{F}\mathbf{W}||_F}$;
**step 4.** Return $\mathbf{F}, \mathbf{W}$.

---

In the following, we exploit the solution of (18) to handle the three steps discussed above.

### A. Step 1

In this step, the hybrid beamforming for the destination is designed. In this regard, the hybrid beamforming matrices of the $k$th user can be optimized by minimizing the third term of (15).

$$\min_{\mathbf{F}_{dk},\mathbf{W}_{dk}} \quad ||\mathbf{F}_{dk}\mathbf{W}_{dk} - \tilde{\mathbf{W}}_{dk}||_F^2$$
$$\text{s.t.} \quad |\mathbf{F}_{dk}(i,j)|^2 = 1. \quad (19)$$

We can solve this problem (19) by applying the MP algorithm in Table I while ignoring step 3. This solution, however, yields very large objective function which consequently decrease the achieved sum rate of the system especially at high signal to noise ratio (SNR) regions. For this reason, we introduce the following constraint

$$\text{trace}(\mathbf{F}_{dk}\mathbf{W}_{dk}\mathbf{W}_{dk}^H\mathbf{F}_{dk}^H) = \text{trace}(\tilde{\mathbf{W}}_{dk}^H\tilde{\mathbf{W}}_{dk}). \quad (20)$$

For the $k$th destination's hybrid beamforming optimization problem (19) with constraint (20), we can apply the MP algorithm in Table I. For this problem, the initialization parameters are: $\mathbf{W}_{res} = \mathbf{W}_{opt} = \tilde{\mathbf{W}}_{dk}$, $\mathbf{D}_1 = \mathbf{I}$, $P = \text{trace}(\tilde{\mathbf{W}}_{dk}^H\tilde{\mathbf{W}}_{dk})$.

### B. Step 2

In this step, we optimize $\mathbf{F}_{r1}, \mathbf{F}_{r2}$, and $\mathbf{W}_r$ for the fixed $\mathbf{W}_d$ and $\mathbf{F}_d$ obtained from Step 1. These variables can be optimized by considering the second term of (16) and can be obtained by solving the following problem

$$\min_{\mathbf{F}_{r1},\mathbf{W}_r,\mathbf{F}_{r2}} \quad ||\mathbf{B}_r\mathbf{F}_{r1}\mathbf{W}_r\mathbf{F}_{r2} - \mathbf{A}_2||_F^2$$
$$\text{s.t.} \quad \text{trace}(\mathbf{F}_{r1}\mathbf{W}_r\mathbf{F}_{r2}(\mathbf{C}_r\mathbf{C}_r^H + \sigma_r^2\mathbf{I})\mathbf{F}_{r2}^H\mathbf{W}_r^H\mathbf{F}_{r1}^H) = P_r,$$
$$|\mathbf{F}_{r1}(i,j)|^2 = 1, \quad |\mathbf{F}_{r2}(i,j)|^2 = 1, \quad (21)$$

where $\mathbf{B}_r = \mathbf{W}_d^H \mathbf{F}_d^H \mathbf{H}^H$ and $\mathbf{C}_r = \mathbf{G}\tilde{\mathbf{W}}_t$.

As can be seen from (21), there are two analog beamforming matrices and one digital beamforming matrix at the relay station. For this reason, we propose a two-step MP algorithm to solve this problem. First, by assuming $\hat{\mathbf{W}}_r = \mathbf{W}_r\mathbf{F}_{r2}$, we apply the MP algorithm to solve the problem with variables $\mathbf{F}_{r1}$ and $\hat{\mathbf{W}}_r$. Then, we optimize $\mathbf{W}_r$ and $\mathbf{F}_{r2}$ by employing $\hat{\mathbf{W}}_r$ as

$$\min_{\mathbf{W}_r,\mathbf{F}_{r2}} \quad ||\mathbf{F}_{r2}^H\mathbf{W}_r^H - \hat{\mathbf{W}}_r^H||_F^2$$
$$\text{s.t.} \quad \text{trace}(\mathbf{F}_{r1}\mathbf{W}_r\mathbf{F}_{r2}(\mathbf{C}_r\mathbf{C}_r^H + \sigma_r^2\mathbf{I})\mathbf{F}_{r2}^H\mathbf{W}_r^H\mathbf{F}_{r1}^H) = P_r,$$
$$|\mathbf{F}_{r2}(i,j)|^2 = 1. \quad (22)$$

According to (21) and (22), the MP algorithm for hybrid beamforming at the relay station is presented in Table II:

TABLE II
RELAY HYBRID BEAMFORMING

---

**step 1.** Let $\mathbf{F} = [\ ]$, $\mathbf{D}_1 = \mathbf{B}_r$, $\mathbf{W}_{res} = \mathbf{W}_{opt} = \mathbf{A}_2$;
**step 2.** For $i = 1 : N_r$ do
  Apply **step 2** in Table I
  end;
**step 3.** Scale $\mathbf{W} = \sqrt{P_r}\frac{\mathbf{W}}{||\mathbf{F}\mathbf{W}(\mathbf{C}_r+\sigma\mathbf{I})||_F}$;
**step 4.** Return $\mathbf{F}_{r1} = \mathbf{F}$, $\hat{\mathbf{W}}_r = \mathbf{W}$.
**step 5.** Let $\mathbf{F} = $ Empty Matrix, $\mathbf{D}_1 = \mathbf{I}$, $\mathbf{W}_{res} = \mathbf{W}_{opt} = \hat{\mathbf{W}}_r^H$;
**step 6.** For $i = 1 : N_r$ do
  Apply **step 2** in Table I
  end;
**step 7.** Scale $\mathbf{W} = \sqrt{P_r}\frac{\mathbf{W}}{||\mathbf{F}_{r1}\mathbf{W}^H\mathbf{F}^H(\mathbf{C}_r+\sigma\mathbf{I})||_F}$;
**step 8.** Return $\mathbf{F}_{r2} = \mathbf{F}^H$, $\mathbf{W}_r = \mathbf{W}^H$.

---

## C. Step 3

For the given $\mathbf{F}_d$, $\mathbf{W}_d$, $\mathbf{F}_{r1}$, $\mathbf{W}_r$, and $\mathbf{F}_{r2}$, the problem (17) can be formulated as a source hybrid beamforming optimization problem, which can be expressed as

$$\min_{\mathbf{F}_t, \mathbf{W}_t} \quad ||\mathbf{B}_t \mathbf{F}_t \mathbf{W}_t - \mathbf{A}_1||_F^2$$
$$\text{s.t.} \quad \text{trace}(\mathbf{W}_t^H \mathbf{F}_t^H \mathbf{F}_t \mathbf{W}_t) = P_s$$
$$|\mathbf{F}_t(i,j)|^2 = 1 \quad (23)$$

where $\mathbf{B}_t = \mathbf{W}_d^H \mathbf{F}_d^H \mathbf{H}^H \mathbf{F}_{r1} \mathbf{W}_r \mathbf{F}_{r2} \mathbf{G}$, and $\mathbf{A}_1$ is defined in (16). This problem can also be solved by employing the MP algorithm in Table I which is similar to Step 1 above. The initialization parameters for the source optimization problem are: $\mathbf{W}_{res} = \mathbf{W}_{opt} = \mathbf{A}_1$, $\mathbf{D}_1 = \mathbf{B}_t$, $P = P_s$.

## VI. SIMULATION RESULTS

This section presents simulation results of the proposed hybrid beamforming design. We assume a mm-wave system operating at 28 GHz where the source, relay and each destination have equipped with multiple antennas and limited number of RF chains. All the simulation results are generated by averaging 5000 randomly chosen channel realizations. The simulation results are presented for different parameters, including the signal to noise ratio (SNR) of the whole system and numbers of RF chains and paths at the source, relay and destinations. We also examine the effects of AoA and AoD estimation errors on the sum rate of the hybrid beamforming designs. For simplicity we assume that the total power constraints at the source and destination are equal. We have used $M_s = 128$, $M_r = 32$, $M_{dk} = 32$ $d = 0.5\lambda$, $K = 4$, $L_g = 16$, $L_{hk} = 4, \forall k$ and $P_s = P_r$. The SNR which is defined as $SNR = \frac{P_{av}}{\sigma^2}$ is controlled by varying $P_{av} = \frac{P_s}{K}$ while keeping the noise power to unity. For the simulation, we assume that the system employs a scheduling mechanism such that destinations of having almost the same path loss are served simultaneously. For this reason, we assume that each of the destinations has the same path loss.

### A. Comparison of digital and hybrid beamformings

In this simulation, we compare the sum rate obtained by utilizing the digital and hybrid beamforming designs. In this regard, we assume that the system employs an approach to estimate the AoA and AoD's of all channel vectors. In this regard, we assume that the AoA and AoD's are estimated with estimation accuracy error of $0.02\%$. The number of RF chains at the source $N_s$, relay $N_r$ and each destination $N_{dk}$ are set to $N_s = N_r = 16$ and $N_{dk} = 4, \forall k$. Fig. 2 shows the achievable rates of the digital and hybrid beamformings for these settings. As can be seen from this figure, the proposed hybrid beamforming scheme achieves almost the same performance as that of the digital beamforming.

### B. Effect of Number of RF chains

This subsection discusses the simulation results obtained from examining the joint effects of number of RF chains and paths on the performances of the proposed hybrid beamforming designs. To this end, we vary the number of RF chains at the source, relay and destinations. The number of paths $L_g$ in $\mathbf{G}$ is also varied following the number as RF chains at source which means $L_g = N_s$. As can be seen from Fig. 3, when $N_s$ is varied from 16 to 24, $N_r = N_s$ and $N_d$ is varied from 4 to 12, the achievable sum rate of the proposed hybrid beamforming design increases as the numbers of RF chains at the source, relay and destinations increase which is the expected result.

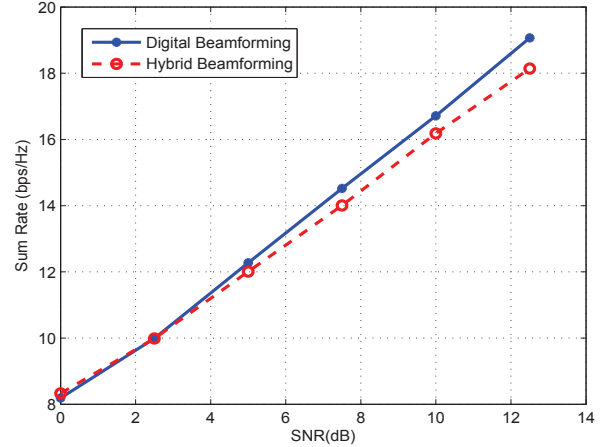

Fig. 2. Digital beamforming, hybrid beamforming with different SNR

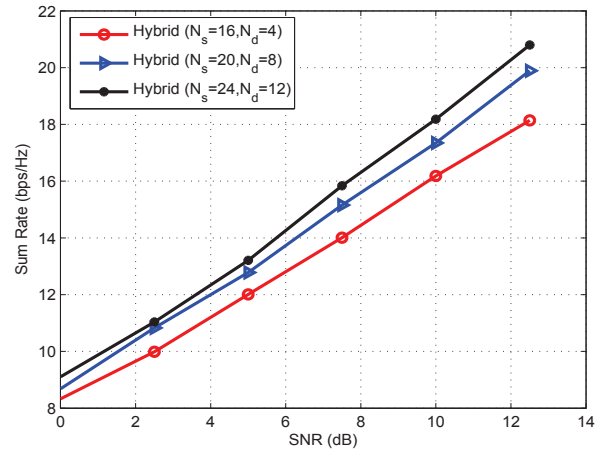

Fig. 3. Different RF chains at source, relay and all destinations

### C. Effect of the AoA and AoD estimation errors

In this simulation we examine the effect of AoA and AoD estimation errors of the proposed hybrid beamforming algorithm which is presented in Fig. 4. As can be seen from this figure, the achievable rate decreases as the AoA (AoD) estimation errors increase. From this figure, one can also notice that the MP based hybrid beamforming is very sensitive to such errors which is unavoidable in practice.

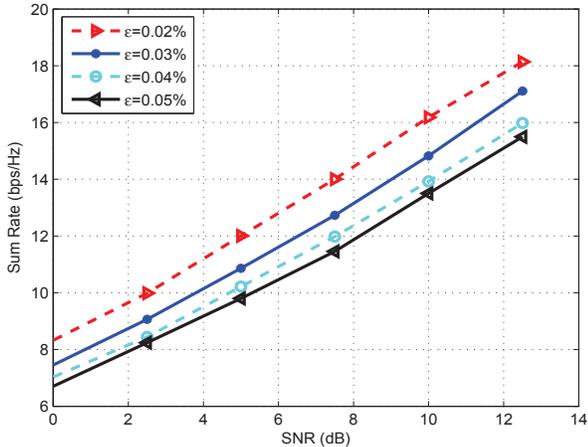

Fig. 4. Different channel estimation error

## VII. CONCLUSION

This paper proposed new hybrid, analog-digital, beamforming for a multiuser mm-wave relay system for the sum rate maximization problem. The proposed hybrid beamforming is designed indirectly by considering MSE based problem while utilizing the solution of digital beamforming. To this end, we assume that the digital beamforming utilizes the well-known BD approach. Under this assumption, the considered problem is solved as follows: First, the sum rate maximization problem is formulated as the minimization of the MSE between the received signal of the hybrid and digital beamforming designs. Then, the hybrid beamforming matrices of the source, relay and each destination is designed by leveraging compressive sensing techniques. Simulation results validate that the proposed hybrid beamforming design achieves close performance to that of the digital one. Furthermore, the effects of the number of RF chains and paths together, and the accuracy of AoA and AoD estimators on the sum rate of the hybrid beamforming designs have been examined . Computer simulations demonstrate that the total sum rate of the hybrid beamforming increases when the number of RF chains and paths increase (or the accuracy of the AoA (AoD) estimator improves).

## VIII. ACKNOWLEDGMENTS

This paper is supported by the following grants: National Science Foundation of China under grant 61372135 and 111 project B08038, and the NSERC Strategic Project Grant (UM Project #40247).